# Efficient Electro-Optical Modulation Based on Indium Tin Oxide

**Kaifeng Shi and Zhaolin Lu**[*]

*Microsystems Engineering, Kate Gleason College of Engineering,
Rochester Institute of Technology, Rochester, New York, 14623, USA
\*Corresponding author: zhaolin.lu@rit.edu*

**Abstract:** We experimentally demonstrate several electro-optical modulators based on transparent conducting oxides. Our previous work demonstrated the modulator structure on glass substrate with broadband bias polarity-dependent modulation. Further exploration shows similar modulation effect of the modulator on quartz and silicon substrate.

**Keywords:** Electro-optical Modulator, Indium Tin Oxide (ITO), Broadband.

## 1. Introduction

As one of the most critical devices in optoelectronic integrated circuits, the electro-optical (EO) modulator has long been suffering from the poor EO properties of conventional materials, even for some well-known EO material, such as lithium niobate [1,2], which inhibit significant modulation within a compact modulator. The length of some phase modulators is on the order of millimeters [3]. In recent years, many EO modulators have been proposed on the purpose of small footprint and better performance. For instance, silicon resonator modulators enhance the EO effect due to large quality factor of resonant cavity, which can shrink their dimensions to tens of micrometers [4-6]. However, resonator modulators usually suffer from bandwidth limitation, temperature fluctuation as well as fabrication tolerance. On the other hand, non-resonant modulators exhibit broadband performance, but they lose the compactness [7].

Development in plasmonics opens a new vista for EO modulators [8-12]. Some EO modulators utilize a metal-oxide-semiconductor (MOS) structure and show a hybrid plasmonic mode [12,13], thus having the advantages of lower losses and easy integration with CMOS platforms. A recent reported plasmonic phase modulator has a very small length (29μm) and high speed (40Gbit/s), which realizes EO modulation by exploiting Pockel effect in nonlinear polymer [7]. The choice of active EO material is another important concern that affects device performance. In addition to lithium nibobate, silicon and polymers, graphene has also been reported as an active material for nanoscale EO modulator working around telecommunication wavelength [14].

Recently, the family of transparent conducting oxides (TCOs) has played an important role in novel EO modulators. TCOs are doped metal oxides. They have a large bandgap which make them visibly transparent. They can be heavily doped to exhibit high electrical conductivity, and more importantly their plasma frequencies lie in the near-infrared (NIR) regime. The exploration of TCOs as the plasmonic metamaterial for NIR applications can be traced back to decades ago [15]. Some comparative studies have been reported [16-19], showing their advantages of low loss and fabrication compatibility. Indium tin oxide (ITO), a well-known representative of TCOs, has been widely used as transparent electrodes in solar cells and display [20-22]. The carrier concentration in TCOs can be controlled by manipulating the concentration of oxygen vacancies and interstitial metal dopants, which result in optical properties change.

The optical dielectric constant of TCOs can be approximated by the Drude model,

$$\varepsilon = \varepsilon' + j\varepsilon'' = \varepsilon_\infty \left[ 1 - \frac{\omega_p^2}{\omega(\omega + j\gamma)} \right] \quad (1)$$

where $\varepsilon_\infty$ is high frequency dielectric constant, $\omega$ is the angular frequency of the light wave, $\gamma$ is the damping coefficient of free carriers related to optical losses. The plasma frequency is defined by

$$\omega_p = \sqrt{\frac{Ne^2}{\varepsilon_\infty \varepsilon_0 m^*}} \quad (2)$$

which depends on carrier concentration $N$ and the electron effective mass $m^*$. According to the above equations, the optical material's dielectric constant shifts with different carrier concentrations. In Ref. 23, unity-order index change of ITO film in a MOS structure is reported by voltage-induced accumulation carriers at ITO-insulator interface. Therefore, by engineering the carrier concentrations in TCOs, the absorption of active material, which is directly related to its index, or dielectric constant, could be tuned. TCOs are now widely used in electro-absorption (EA) modulators [24-26], but only few experiments are reported.

## 2. Broadband multilayer EO modulator

In our previous work, we have investigated an EO modulator configuration based on ~25nm thick ITO (as the active medium) and ~5μm thick electrolyte gel (as gate "oxide"), where we observed the bias polarity-dependent modulation effect: the absorption of the EO modulator could be increased by negative bias or decreased by positive bias [27]. Furthermore, in order to make more efficient modulators, we apply high-k dielectric material in our designed EO modulator structure, as illustrated in Fig. 1(a) [28]. Here high-k material $HfO_2$ is utilized as the gate oxide for its ultrahigh permittivity as well as process stability [29]. Aluminum is chosen as the metal for its excellent conductivity, low absorption in the NIR regime, as well as low cost. Another advantage of using aluminum layer is that light absorption can be directly measured by $1-R$ ($R$ is the power reflectance) in the device.

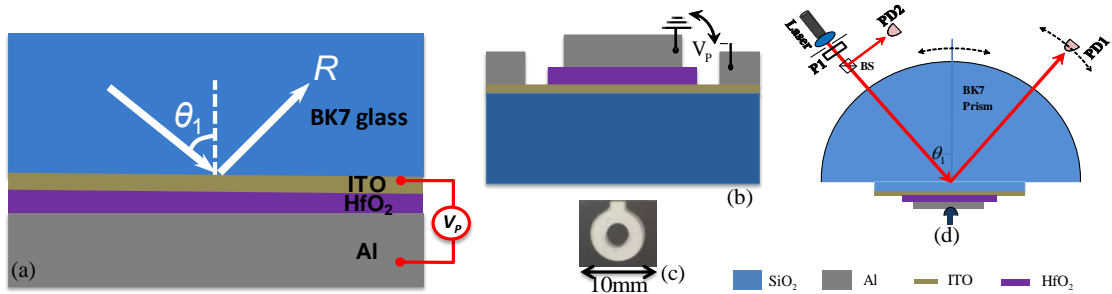

Figure 1. (a) Illustration of ITO-based multilayer EO modulator. (b) Cross section of the proposed multilayer ITO modulator. (c) Picture of the fabricated modulator. (d) Illustration of the setup for ATR measurement.

The cross section of the ITO-based EO modulator is shown in Fig. 1(b). The fabrication of the modulators starts from ITO film deposition on transparent glass, by the method of physical vapor deposition (PVD) process. The thickness of ITO film is measured in the range of 10~12nm. After post annealing process, the sheet resistance of ITO film is measured around 400~600Ω/□. After that, 50nm thick $HfO_2$ is deposited by atomic layer deposition (ALD). Finally, a 200nm thick layer of aluminum is deposited on top by E-beam evaporation. The picture of the fabricated sample is shown in Fig. 1(c).

We built attenuated total reflectance (ATR) setup (Fig. 1(d)) to measure the light reflectance of the modulators by sweeping the incident angle [30, 31]. In the experiments, we first measured the reflectance of ITO modulator at different wavelengths ranging from $\lambda_0$=1260nm~1620nm without externally applied voltage.

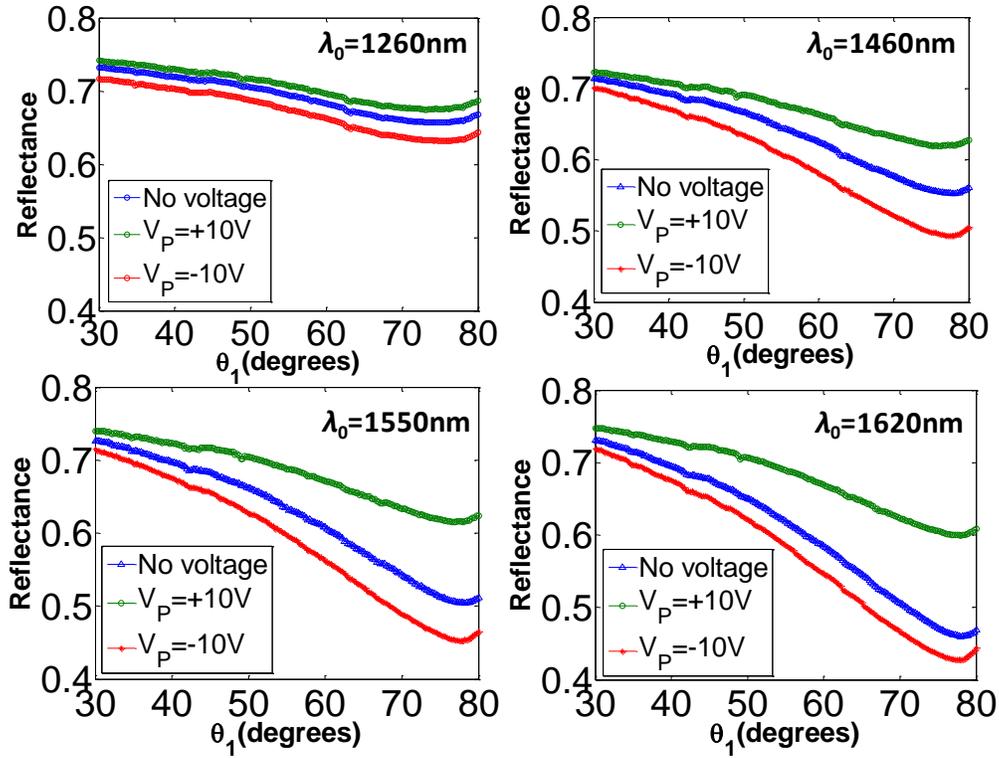

Figure 2. Reflectance as a function of incident angle for the modulator under different applied voltages for different wavelengths.

After that, external voltage $V_P$ was applied. During the experiment, we found that the reflectance (or 1−absorption) from the modulator could either be increased or decreased by applying positive or negative voltage, respectively. We conducted measurements with an externally applied voltage of different polarities but the same magnitude $\pm V_P$. For different wavelengths, the measured reflectance of the modulator with different applied voltages, as a function of $\theta_1$ with a TM-polarized incident light beam is shown in Fig. 2(a-d), respectively. The modulation depth, $M(\theta_1)$, as a function of angle $\theta_1$ at a given wavelength can be defined as: $M(\theta_1) = |R_{+Vp} - R_{-Vp}|/R_0$, where $R_0$ is the experimentally measured reflectance without applied voltage; $|R_{+Vp} - R_{-Vp}|$ is the magnitude of the difference of the reflectance under applied voltages. From Fig. 3, broadband EO modulation has been achieved and the largest modulation depth obtained at a specific angle, $\theta_1=78°$, is $M_{max}(78°)=37.42\%$ for $\lambda_0=1620nm$.

We attribute the modulation mainly to the change of the free carrier concentration in the voltage induced active layer in ITO at the interface. For instance, when negative bias is applied, excess positive carriers will be induced at the ITO-$HfO_2$ interface, which results in major carrier (holes) accumulation in the active ITO layer; on the contrary, the active layer will be depleted under positive bias.

According to our measurement under applied DC voltage, the modulation speed of the modulator is quite low. It might need one minute or more to let most of the absorption modulation complete. The relatively fast modulation (within milliseconds) only witnesses a very small modulation depth.

## 3. Faster modulator on quartz or silicon substrate

More recently, for the purpose of improving the working speed of our EO modulator as well as its easier integration with the current CMOS platforms, we examined the performance of the modulator on quartz and silicon substrate, respectively.

We deposited ITO film by PVD75 sputtering tool [32] where the substrate could be heated to as high as 350 ℃ to create denser films, so the post annealing process is no longer needed. The oxygen flow ratio could be adjusted to control the uniformity and conductivity of the ITO film. We chose 200 ℃ for substrate heating with an oxygen flow ratio of 2% in the deposition process, and a 70-second deposition time yield 12~14nm thick ITO film. The sheet resistance of the ITO film on quartz substrate was measured to be 200~350 Ω/□. Considering that a larger gate-source electric field may result in better modulation performance, we decreased the thickness of $HfO_2$ to 10nm in the ALD process. The thickness of the Al film remains to be 200nm, and the samples keep the same shape and dimension as Fig. 2(b).

We tested the fabricated samples in the same setup as shown in Fig. 2(c). The modulation angle, where the modulation depth reached its maximum, was first identified by the $R$-$\theta_1$ measurement under DC bias voltages. After that, we fixed the sample at the modulation angle, and imposed AC electric signal on the modulator. Due to the speed limit of the Ge photodiode, the modulation depth measurement was conducted at the frequency of 1 kHz. For $\lambda_0$=1260nm incident light, the power reflected by the modulator on quartz substrate under different peak-to-peak AC square signals are plotted in Fig. 3. The percentage in the bracket shows the maximum modulation depth under the applied voltage. Similar to previous result, a larger electric signal produces a higher modulation depth with a near-linear relation.

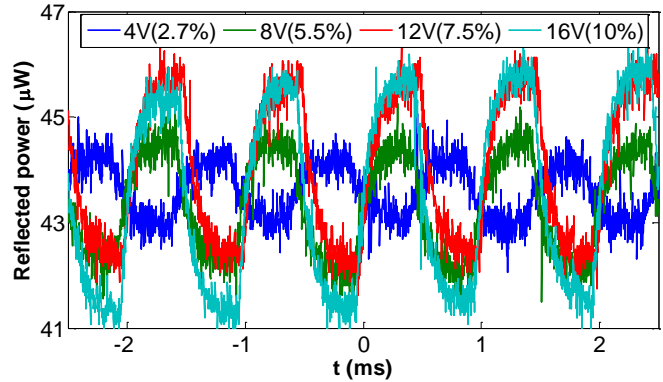

Figure 3. Reflected power of EO modulator on quartz as a function of time at the modulation angle for $\lambda_0$=1260nm.

The breakdown voltage for the ITO EO modulator on quartz is about ±8V, which is smaller compared to the previous test. Note that a thinner gate oxide layer results in a larger electric field under the same applied voltage. Since the substrate is not conductive, there generates much heat within the thin films due to a large electric current, and finally the materials get burned.

EO modulators on highly doped silicon substrate show similar modulation effect, but they do not breakdown, even under a much larger electric signal. A reasonable explanation might be: current could go through the substrate and heat gets well dissipated. However, there is a tradeoff: the maximum modulation depth observed under a 20V peak to peak AC signal (equivalently ±10V at AC frequency) is less than 5%.

In order to well identify the working speed of the EO modulators on quartz and doped silicon, we utilized an amplified InGaAs photodetector, which has an AC coupling characteristic (i.e. the DC component is removed, and only the power fluctuation is recorded). Two types of doped silicon with different doping levels were used. After ITO deposition, they exhibit sheet resistances of 120~150 Ω/□ and ~2 Ω/□, respectively.

Figure 4(a) illustrates the oscilloscope measurement of reflected power fluctuation of EO modulator on quartz under electric signals of different frequencies. The signals are still square waves. As can be seen, the reflected power keeps square shape with similar magnitude when the signal frequency is 10 kHz or 20 kHz; while for 50 kHz signal, the rising and falling edges become sharper and the magnitude decreases a little. The change is more significant at 100 kHz. Almost similar effect shows up for modulator on doped silicon with higher sheet resistance. However, for modulator on highly doped silicon with a much lower sheet resistance, we can observe modulation at a frequency up to MHz. Fig. 4(b) depicts the EO modulation under 30V peak to peak, 0.5MHz AC signal. Obviously, the working speed of the modulator has been improved due to a smaller RC delay, which results from a less resistive substrate.

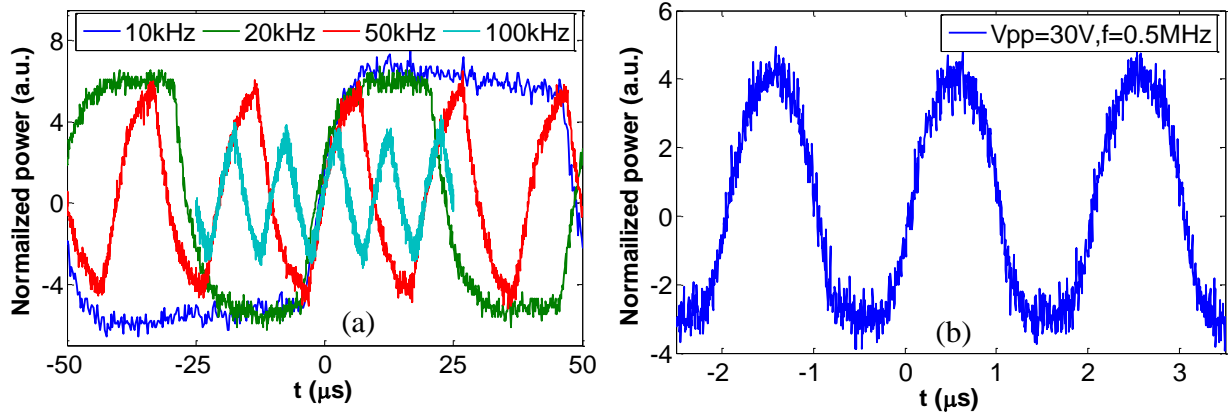

Figure 4. Oscilloscope measurement of reflected power fluctuations of EO modulator on (a) quartz as a function of time at the modulation angle for $\lambda_0$=1260nm and (b) highly doped silicon as a function of time at the modulation angle for $\lambda_0$=1280nm.

Compared to the EO modulator in Section 2, the newly explored ones on quartz and doped silicon can work at a much higher speed of tens of kHz; however, it is still too small for optoelectronic applications, which may need modulators to work at tens or hundreds of GHz. We believe that RC delay is the main factor that limits the speed of the device. We have shown that decreasing resistance could lead to a higher speed, and we are trying to optimize this by further ion implantation in the doped silicon substrate. We are also expecting a speed boost of several orders by shrinking the size of the modulator from millimeter scale to micrometer scale.

Despite the improved speed, the working wavelength range of the modulators in this section has become smaller. The modulation effect occurs at wavelengths between of 1260nm and 1440nm (our laser sources are limited within 1260~1620nm), and the maximum modulation depth is quite sensitive to wavelength. We owe this mainly to the property change of ITO due to different fabrication process.

## 4. Conclusion

To summarize, we have experimentally demonstrated multilayer EO modulators based on ITO. Our former modulator structure on glass substrate has achieved broadband EO modulation with a relatively large modulation depth. We further optimized the design and explored the modulator built on quartz and silicon substrate, respectively. We are still working on the optimization and trying to make the modulator into micrometer scale, aiming at faster and more efficient performance. The successful illustration of the modulator on silicon substrate will make it more practical in on-chip integration and applications.

**Acknowledgement**

The publication was made possible by Grant Number W911NF-12-1-0451 from the United States Army. This material is based upon work partially supported by the National Science Foundation under Award No. ECCS-1308197.